\documentclass[epj]{svjour}
\usepackage{latexsym,amsmath}
\usepackage{graphics,dsfont}
\newcommand{\s}{s^2}
\begin{document}
\title{Fluctuation symmetry in a two-state Markov model.}
\author{Tim Willaert \and Bart Cleuren \and Christian Van den Broeck}                     
\institute{Hasselt University, 3590 Diepenbeek, Belgium}
\date{Received: date / Revised version: date}
%
\abstract{
We show that the scaled cumulant generating and large deviation function, associated to a two-state Markov process involving two processes, obey a symmetry relation reminiscent of the fluctuation theorem, independent from any conditions on the transition rates.  The Legendre transform leading from the scaled cumulant generating function to the large deviation function is performed in an ingenious way, avoiding the sign problem associated to taking a square root. Applications to the theory of random walks and to the stochastic thermodynamics for a quantum dot are presented.
} 
\maketitle
\section{Introduction}
\label{intro}
The so-called fluctuation theorem first appeared in studies of thermostated systems in the 
early 1990 ies, and has since been rederived in many different settings. In one of its forms, it prescribes a constraint on the large deviations in the total amount of heat  $Q$ transfered between two reservoirs (temperatures $T_1$ and $T_2$) over a time $t$, by a device with bounded energy. More precisely, 
the scaled cumulant generating function $f(\gamma)=\lim_{t\rightarrow \infty}\ln\langle \exp{\gamma Q}\rangle/t$, obeys the relation $f(\gamma)=f(\gamma_0-\gamma)$, with  
$\gamma_0=1/ T_2-1/ T_1$ the thermodynamic force associated to the heat flux between the reservoirs (and we used the convention $k_B=1$).  This fluctuation theorem and its variants all rest on physical assumptions imposed on the parameters of the model, for example the local detailed balance conditions obeyed by the transition rates in Markovian descriptions.
In the present paper, we show that a similar symmetry property holds in a two-state Markov process undergoing two exchange processes. The remarkable fact is that this relation is obeyed without any conditions on the transition rates and independent of any thermodynamic interpretation. It breaks down when more processes or more states are considered. It reduces to the usual fluctuation theorem when the proper conditions for a thermodynamic interpretation are valid (i.e., when the rates obey a local detailed balance conditions).
\section{Two-state Markov process undergoing two exchange processes}
\begin{figure}
\begin{center}\resizebox{0.22\textwidth}{!}{
	\includegraphics{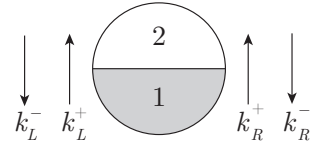}}
	\end{center}
\caption{Two-state model. For every clockwise rotation, i.e. a sequence of transitions $1 \rightarrow 2$ via mechanism $L$ followed by $2 \rightarrow 1$ via mechanism $R$, $x$ changes by an amount $\delta_L-\delta_R$.}
\label{fig:2ss}
\end{figure}

\begin{figure*}
\begin{center}
\resizebox{0.40\textwidth}{!}{\includegraphics{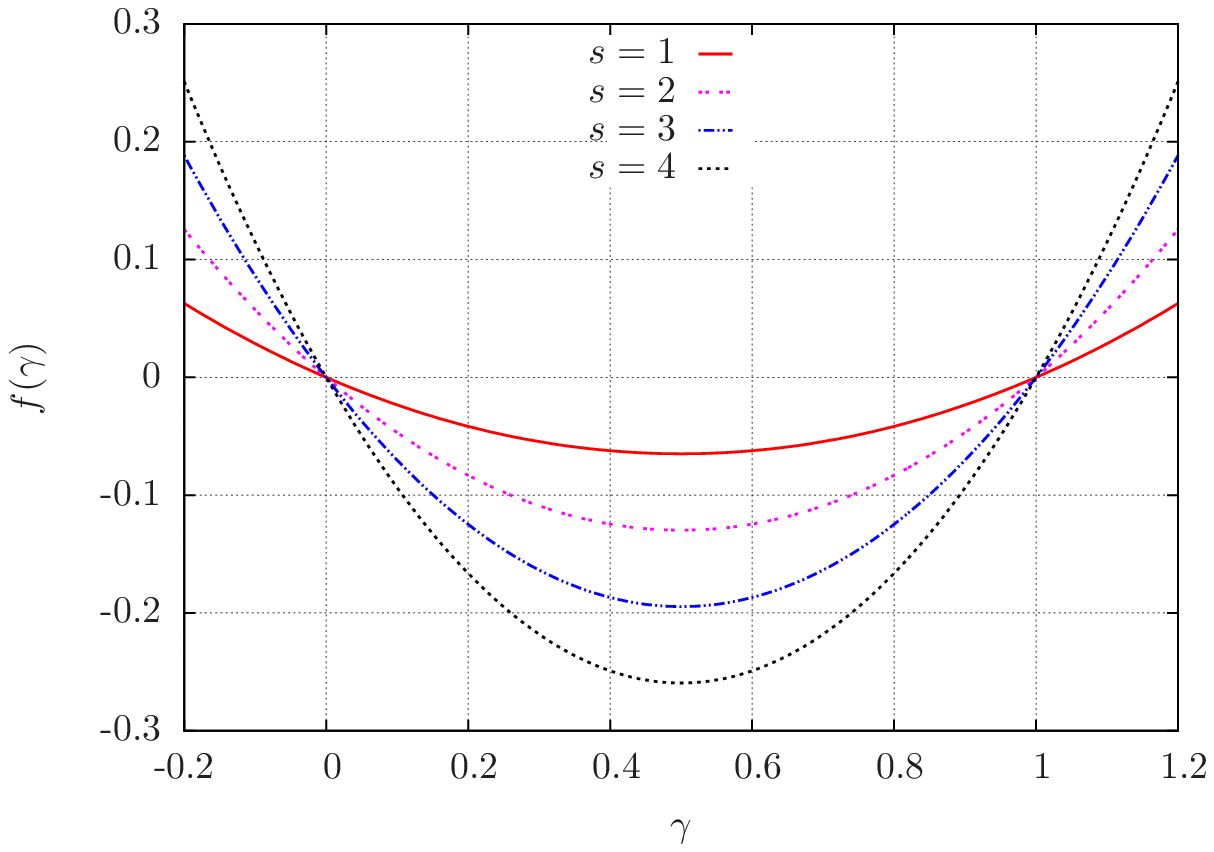}}\hspace{1.2cm}
\resizebox{0.40\textwidth}{!}{\includegraphics{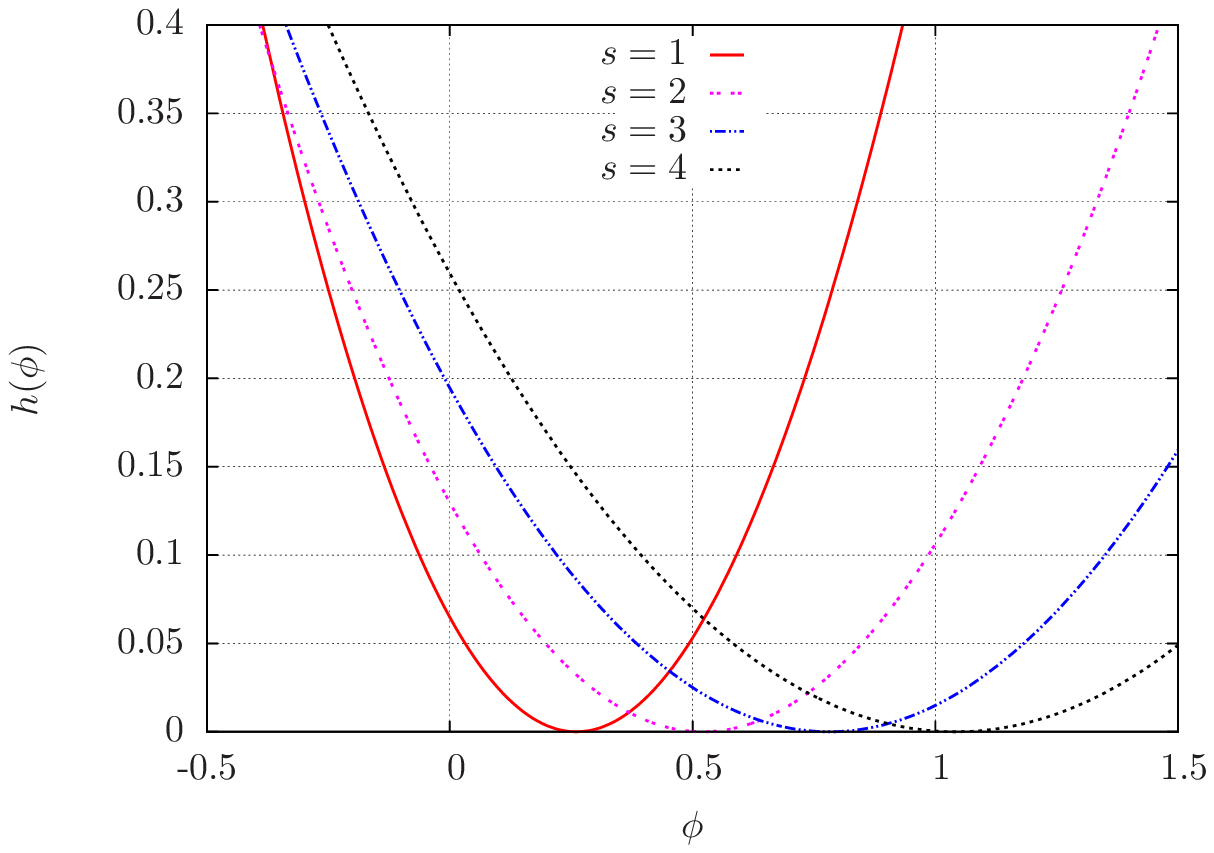}}
\end{center}
\caption{Graph of the CGF $f(\gamma)$ (left) and the corresponding LDF $h(\phi)$ (right). The curves are obtained by varying $k$ and $s$ for fixed ratio $k/s=2$. In all cases $\gamma_0=1$. The zero minimum of the LDF is reached at the average value given by (\ref{1cum}).}
\label{figCGF}
\end{figure*}

We consider a dichotomic Markov process, in which the transitions between the two states $i$, $i=1$ and $i=2$, are induced by two different mechanisms,  refered to as $L$(eft) and $R$(ight). We denote the transitions rate $k^{\pm}_{\nu}$ for going from state 1 to state 2 ($+$) and from state 2 to 1 ($-$), due to mechanism $\nu \in \{L,R\}$, respectively, cf. Figure \ref{fig:2ss} for a schematic representation. 
We will be interested in a quantity that undergoes a change with every transition, more precisely it changes by an amount  $\pm \delta_\nu$  for a  transition from $1$ to $2$ (+ sign) and from $2$ to $1$ (-sign) due to mechanism $\nu$. Let $x$ denote the cumulated value of this quantity during the time interval from $[0,t]$. 
Since $x$ depends in a deterministic way on the stochastic process $i$,  the joint set $x,i$ still defines a Markov process. The corresponding probability  $\vec{p}(x;t)=\{p_i(x;t), i=1,2\}$ obeys the following Master equation:
\begin{equation}\label{mex}
\dot{\vec{p}}(x;t)=\mathds{W}\vec{p}(x;t)
\end{equation}
where the  matrix $\mathds{W}$ is given by:
\begin{equation}
\mathds{W}=
\left( \begin{array}{cc}
-k^{+}_L-k^{+}_R &  k^{-}_L e^{\delta_L \partial_x}+k^{-}_R e^{\delta_R \partial_x}\\
k^{+}_L e^{-\delta_L \partial_x}+k^{+}_R e^{-\delta_R \partial_x}& -k^{-}_L-k^{-}_R
\end{array}\right).
\end{equation}
The meaning of $x$ can be visualized by introducing the dimensionless quantity $n$ defined through $x\equiv n(\delta_L-\delta_R)$. Since $\delta=\delta_L-\delta_R$ equals the change of $x$ during a full clockwise rotation, cf. Figure \ref{fig:2ss}, an integer value of $n$ can be interpreted as the net number of such clockwise rotations. To proceed, it is convenient to switch to the following generating function:
\begin{equation}
\vec{g}(\gamma,t)=\int dx e^{-\gamma n} \vec{p}(x;t)  \label{CGF}
\end{equation}
Its evolution equation follows immediately from the above master equation (\ref{mex}):
\begin{equation}
\dot{\vec{g}}(\gamma,t)=\mathds{M}(\gamma)\ \vec{g}(\gamma,t)
\end{equation}
with
\begin{equation}\label{M}
\mathds{M}=
\left( \begin{array}{cc}
-k^{+}_L-k^{+}_R &  k^{-}_Le^{\gamma\frac{\delta_{L}}{\delta}}+k^{-}_Re^{\gamma\frac{\delta_{R}}{\delta}} \\
{k^{+}_L}{e^{-\gamma\frac{\delta_{L}}{\delta}}}+{k^{+}_R}{e^{-\gamma\frac{\delta_{R}}{\delta}}} & -k^{-}_L-k^{-}_R
\end{array}\right)
\end{equation} 
The asymptotic properties of $n$ in the longtime limit $t \rightarrow \infty$ can now be obtained as follows. Consider the scaled cumulant generating function (CGF) $f(\gamma)$ defined by
\begin{equation}
f(\gamma)=\lim_{t\to\infty} \frac{1}{t} \ln\left\langle e^{-\gamma n} \right\rangle=\lim_{t\to\infty} \frac{1}{t} \ln\left[\sum_i g_i(\gamma,t)\right].
\end{equation}
The vector $\vec{g}$ evolves  over a time duration $t$ according to the propagator $e^{\mathds{M}.t}$, hence $f$ is just the largest eigenvalue of the matrix $\mathds{M}$. 
Before proceeding to an explicit analytic solution, we make the following observation. The eigenvalue equation of the matrix $\mathds{M}$ is given by: 
\begin{equation}\label{eve}
\lambda^2-\mathrm{Tr}\;\mathds{M}\;\lambda+ \det\;\mathds{M}=0.
\end{equation}
From (\ref{M}), one readily verifies that $\mathrm{Tr}\;\mathds{M}$ is  independent of  $\gamma$, while $\det\;\mathds{M}$ is invariant under the transformation $\gamma \rightarrow \gamma_0-\gamma$,   with 
\begin{eqnarray}\label{gamma0}
\gamma_0=\log\left(\frac{k_L^+ k_R^-}{k_L^- k_R^+}\right) \label{X}.
\end{eqnarray}
As a result the eigenvalues and in particular the largest eigenvalue will inherit this symmetry property, hence:
\begin{equation}\label{sym_f}
f(\gamma)=f(\gamma_0-\gamma).
\end{equation}
This result can also be derived by noting the existence of the following similarity transformation, see also \cite{Lacoste}:
\begin{equation}
\mathds{M}^{\dagger}(\gamma_0-\gamma)=\mathds{Q} \mathds{M}(\gamma)\mathds{Q}^{-1}
\end{equation}
with
\begin{equation}
\mathds{Q}=\left(\begin{array}{cc}
\sqrt{k^+_L/k^-_L}e^{-\frac{\gamma_0 \delta_L}{2\delta}} &0 \\
0 & \sqrt{k^-_R/k^+_R}e^{\frac{\gamma_0 \delta_R}{2\delta}}
\end{array}\right)
\end{equation}
The above property is formally identical to the above mentioned fluctuation symmetry, while $\gamma_0$ has indeed a structure reminiscent of a thermodynamic force. The sign of $\gamma_0$, positive or negative, determines whether rotations, clockwise or counterclockwise, are favoured. We however stress that the above property is derived without any conditions on the transition rates or any physical interpretation or condition on the quantity $x$.
We proceed with a further analysis of the largest eigenvalue of (\ref{eve}). It is given by
\begin{equation}
f(\gamma)=(\mathrm{Tr}\;\mathds{M})/2 + \sqrt{(\mathrm{Tr}\;\mathds{M})^2 /4-\det\;\mathds{M}},
\end{equation}
which can be rewritten as follows:
\begin{equation}\label{cgfform}
f(\gamma)=k+\sqrt{r^2+\s q(\gamma)}
\end{equation}
with:
\begin{eqnarray}
&&k=(\mathrm{Tr}\;\mathds{M})/2 \;\;\;;\;\;\; \s=\sqrt{k^{+}_Lk^{-}_Lk^{+}_Rk^{-}_R}\\
&&r^2=k^2-k^{+}_Lk^{-}_R-k^{-}_Lk^{+}_R=k^2-\s q(0)>0\\
&&q(\gamma)=\rho(\gamma)+\rho(\gamma)^{-1}\;\;\;\;;\;\;\;\;\rho(\gamma)=e^{\gamma_0/2-\gamma} \label{qrho}
\end{eqnarray}
A graph of (\ref{cgfform}) is shown in Figure \ref{figCGF}. The goal now is to investigate what properties can be derived when the CGF takes the form (\ref{cgfform}) and (\ref{qrho}).\\
While this expression is derived within the context of a generic two-state system, we stress that this same form has been found not only for specific cases of this type (e.g. \cite{Lacoste}), but also for slightly different setups (cf. \cite{TwinElev}. for an example of a 4-state system).\newline \noindent
Having obtained the CGF, it is straightforward to calculate the cumulants, which are determined by the series expansion:
\begin{equation}
f(\gamma)=\lim_{t\rightarrow \infty}\frac{1}{t}\left(\sum_{i=1}^{\infty}\frac{\kappa_i (-\gamma)^{i}}{i!}\right).
\end{equation}
The results for the first two cumulants read:
\begin{equation}
\lim_{t\rightarrow \infty}\frac{\langle n \rangle}{t}=\frac{\s \sinh(\gamma_0/2)}{\sqrt{r^2+2\s\cosh(\gamma_0/2)}}=\frac{\s \sinh(\gamma_0/2)}{-k}\label{1cum}
\end{equation}
and
\begin{equation}
\lim_{t\rightarrow \infty}\frac{\sigma^2_n}{t}= \frac{(r^2+2\s).(r^2-2\s)-k^4}{4k^{3}}.
\end{equation}
The symmetry relation (\ref{sym_f}) is also reflected in the cumulants. For example, for small driving forces $\gamma_0 \ll 1$ we recover a result quite similar to fluctuation-dissipation relation in the sense that the average is the variance times $\gamma_0/2$:
\begin{equation}
\lim_{t\rightarrow \infty}\frac{\langle n \rangle}{t}\approx \frac{\s\gamma_0}{2\sqrt{r^2+2\s}}
\; \mbox{ and }\;
\lim_{t\rightarrow \infty}\frac{\sigma^2_n}{t} \approx \frac{\s}{\sqrt{r^2+2\s}}\nonumber 
\end{equation}

\section{Large Deviation Function}
We consider the probability distribution $P(n/t=\phi,t)$ that the cumulated value $n$ of the quantity exchanged with the system takes a given average value $\phi=n/t$. The Large Deviation Function (LDF) $h(\phi)$ is then defined for large time $t$ by
\begin{equation}
P(n/t=\phi,t)\sim e^{-h(\phi)t}
\end{equation}
The LDF is obtained as the Legendre transform of the scaled CGF $f(\gamma)$ \cite{LDF}:
\begin{equation}
h(\phi) \equiv \sup_{\gamma}\left(-f(\gamma)-\gamma\phi\right) \label{Legtf}
\end{equation}
Hence $h(\phi) = -f(\gamma(\phi))-\phi.\gamma(\phi)$ 
where $\gamma(\phi)$ is determined as the inverse of
\begin{equation}\label{faccent}
\phi=-f'(\gamma)=-\frac{\s q'(\gamma)}{2\sqrt{r^2+\s q(\gamma)}}.
\end{equation}
The most straight-forward approach to invert this relation leads to a piecewise continuous expression of the LDF, and is sketched briefly in the appendix for comparison. Such an expression obscures the fact that the LDF, being the Legendre transform of a continuous strictly convex function, also must be continuous. In the following we show how the particular form of the CGF allows to obtain an unambiguous expression for the LDF. First, it is convenient to introduce the complementary function:
\begin{equation}
\bar{q}(\gamma)=\rho(\gamma)-\rho(\gamma)^{-1}
\end{equation}
The reason for using $\bar{q}$ instead of $q$ is that both $\bar{q}$ and $\phi$ change sign at $\gamma=\gamma_0/2$, which is the minimum of $f(\gamma)$ . Using $\rho(\gamma)$ from (\ref{qrho}) we can replace $q$ and $q'$ in (\ref{faccent}) by
\begin{equation}
q(\gamma)=\sqrt{\bar{q}(\gamma)^2+4} \;\;\; ; \;\;\;\;
q'(\gamma)=-\bar{q}(\gamma) \label{qbar1}
\end{equation}
The result is a one-to-one expression between $\phi$ and $\bar{q}$ which is easily inverted
leading to:
\begin{equation}
\s \bar{q}(\phi)=2\phi\sqrt{r^2+2\phi^2+2\sqrt{(\s)^2+r^2 \phi^2+\phi^4}} \label{qbar2}
\end{equation}
Because all quantities below the square root are strictly positive, $\bar{q}$ is an analytic function in $\phi$ without any discontinuities. The expression $\gamma(\phi)$ follows then immediately, since $\rho(\gamma)=(q+\bar{q})/2$ and $\gamma=\gamma_0/2-\log\rho$: 
\begin{equation}
\gamma(\phi)=\frac{\gamma_0}{2}-\log\left[\frac{\sqrt{\bar{q}(\phi)^2+4}+\bar{q}(\phi)}{2}\right] \label{gammaphi}
\end{equation}
This again is an analytic function in $\phi$, so the LDF is as well. A graph of $h(\phi)$ is shown in Figure \ref{figCGF}. Using the fact that $\bar{q}(-\phi)=-\bar{q}(\phi)$ 
we obtain after simplification the symmetry:
\begin{equation}
\gamma(\phi)+\gamma(-\phi)=\gamma_0.
\end{equation}
Combining this result with the symmetry property of the CGF, cf (\ref{sym_f}), leads to an alternative formulation of the fluctuation theorem:
\begin{equation}\label{sym_h}
h(\phi)-h(-\phi)=-\gamma_0\phi.
\end{equation}
We continue by illustrating these properties for two elementery systems.
\section{Illustrations}
\subsection{Random walk}
\begin{figure}
	\begin{center}
		\resizebox{0.40\textwidth}{!}{\includegraphics{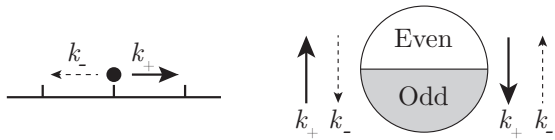}}
	\end{center}
	\caption{(Left) The one dimensional biased random walk. (Right) The mapping onto the two state system.}
	\label{fig:rw}
\end{figure}

The first illustration is provided by the continuous time biased random walk, shown in Figure \ref{fig:rw}. The forward and backward jump rates are $k_+$ and $k_-$ respectively. The quantity $x$ is the position of the walker. A mapping of the random walk to the two state system can be done as follows. Whenever the walker is at an odd (even) integer position, its state is 1 (2). A full clockwise rotation is associated with making two steps forward, hence $k_L^+=k_R^-=v_+$. Similar, a full counter-clockwise rotation is associated with making to steps backward, hence $k_L^-=k_R^+=v_-$ The changes in position correspond to $\delta_L=+1$ and $\delta_R=-1$. For the various parameters we find:
\begin{eqnarray}
k&=&-(k_-+k_+) \quad \quad r^2=2k_-k_+=2\s\\
\gamma_0&=&2\log(k_+/k_-)
\end{eqnarray}
The fact that $r^2=2\s$ simplifies the expressions significantly. For example (\ref{qbar2}) reduces to:
\begin{equation}
\s \bar{q}(\phi)=4\phi\sqrt{\s+\phi^2}
\end{equation}
The resulting CGF is
\begin{equation}
f(\gamma)=-k_+(1-e^{-\gamma/2})-k_-(1-e^{\gamma/2})
\end{equation}
and the LDF reads
\begin{multline}
h(\phi)=k_++k_--2\sqrt{k_+k_-+\phi^2}\\
+2\phi\log\left(\frac{\sqrt{k_+k_-+\phi^2}+\phi}{k_-}\right)
\end{multline}
These expressions are the well known CGF and LDF of a biased random walk cf. \cite{cleuren} (see
 eqs. (27) en (44) therein, $\phi$ plays the role $n/2$). The first two cumulants are
\begin{equation}
\langle n\rangle = \frac{1}{2}\left(k_+-k_-\right) \;\;\;;\;\;\; \sigma^2_n=\frac{1}{4}\left(k_++k_-\right)
\end{equation}
\subsection{Quantum dot: entropy and electron flows}
Another well studied model system is the quantum dot with a fixed single energy level (\cite{Qdot}). The dot is either filled by one electron (state 1) or empty (state 2). Transitions between both states are possible by exchange of an electron with 2 different leads that serve as ideal heat and particle reservoirs $\nu=L,R$, each characterized by a temperature $T^{\nu}$ and chemical potential $\mu_{\nu}$. The transition rates are then expressed in terms of the Fermi distribution $k_{\nu}^{-}=a_{\nu} f_{\nu}$ and $k_{\nu}^{+}=a_{\nu} (1-f_{\nu})$ with $f_\nu=(e^{x_{\nu}}+1)^{-1}$ and $x_{\nu}=\frac{\epsilon-\mu_{\nu}}{T_{\nu}}$ (in units for which $k_B=1$). The constant $a_{\nu}$ $({\nu}=L,R)$ measures the coupling strength between lead ${\nu}$ and the dot. We note that, because of the physical context, the transition rates satisfy the detailed balance condition:
\begin{equation}\label{detbal}
k_{\nu}^{+}/k_{\nu}^{-}=e^{x_{\nu}}
\end{equation} 
For this model we can derive the rate function for various quantities $x$ that correspond to this same clockwise cycling rate. As a first example, consider the particle current from the right lead into the dot.\\
Let $x$ denote the net number of electrons having entered the dot from the right lead during the time interval from $\tau=0$ to $\tau=t$. The change $\delta_\nu$ in this quantity when a transition occurs is then $\delta_{L}=0$ and $\delta_{R}=-1$,
so that $\delta=+1$ and hence $x=n$ for the particle current. Subsituting the values of the transition rates into the expression (\ref{X}) for $\gamma_0$ leads, upon introduction of the notation $x_{\nu}\equiv \log \frac{k^+_{\nu}}{k^-_{\nu}}=\frac{\epsilon-\mu_{\nu}}{T_{\nu}}$ for the argument of the Fermi distribution corresponding to reservoir $\nu$, to:
\begin{equation}
\gamma_0=x_L-x_R=\frac{\epsilon-\mu_L}{T_L}-\frac{\epsilon-\mu_R}{T_R}
\end{equation}
As a second quantity for which the LDF and CGF can be derived from those of $n$, we consider the entropy flow. The changes $\delta_\nu$ for the entropy flow are determined as follows. Suppose an electron moves from the dot to the left reservoir. As a consequence, an amount of heat equal to $\epsilon-\mu_L$ is inserted to the left reservoir and its entropy is raised by an amount $\delta_L=(\epsilon-\mu_L)/T_L=x_L$. A similar reasoning applies for electrons flowing to the right reservoir, and hence $\delta_{\nu}=x_{\nu}$. The conversion factor $\delta$  is given by $\delta=x_L-x_R=\gamma_0$.
\section*{Appendix: Direct Approach}
A direct approach to invert (\ref{faccent}) is to remove the square root in the denominator. By doing so the one-to-one relation between $\phi$ and $\gamma$ is lost. As a simple example, consider the convex function $f(\gamma)=\sqrt{1+\gamma^2}$. (\ref{faccent}) corresponds to 
\begin{equation}
\phi(\gamma)\equiv -f'(\gamma)=-\gamma/\sqrt{1+\gamma^2}
\end{equation}
and a straightforward inversion yields two solutions
\begin{equation}
\gamma^\pm(\phi)=\pm\sqrt{\phi^2/(1-\phi^2)}.
\end{equation}
The Legendre transform is then $h^{\pm}(\phi)=-\sqrt{1+(\gamma^\pm(\phi))^{2}}-\phi.\gamma^{\pm}(\phi)$. Of course, in this example it is clear that $\gamma$ and $\phi$ have opposite sign, so that $h(\phi)$ is indeed continuous. In more elaborate situations, such a simple relation between $\gamma$ and $\phi$ is not immediately obvious.

%
\end{document}